\documentclass[aps,pre,preprint,floatfix,showpacs]{revtex4}
\usepackage{graphicx}
\begin{document}

\title{Passive Random Walkers and River-like Networks on Growing
  Surfaces}

\author{Chen-Shan Chin}

\affiliation{Department of Physics, University of Washington,\\
P.O. Box 351560, Seattle, Washington 98195-1560}

\date{\today}

\begin{abstract}
  Passive random walker dynamics is introduced on a growing surface.
  The walker is designed to drift upward or downward and then follow
  specific topological features, such as hill tops or valley bottoms, of
  the fluctuating surface.  The passive random walker can thus be used
  to directly explore scaling properties of otherwise somewhat hidden
  topological features.  For example, the walker allows us to directly
  measure the dynamical exponent of the underlying growth dynamics.
  We use the Kardar-Parisi-Zhang(KPZ) type surface growth as an
  example.  The word lines of a set of merging passive walkers show
  nontrivial coalescence behaviors and display the river-like network
  structures of surface ridges in space-time.  In other dynamics, like
  Edwards-Wilkinson growth, this does not happen.  The passive random
  walkers in KPZ-type surface growth are closely related to the shock
  waves in the noiseless Burgers equation.  We also briefly discuss
  their relations to the passive scalar dynamics in turbulence.
\end{abstract}


\maketitle

\section{Introduction}
\label{sec:intro}

The research on fluctuations of surfaces during growth has been one of
the major areas of study in nonequilibrium statistical mechanics in
recent decades.  Most research is focused on identifying universality
classes and the scaling behavior of surface morphology, e.g., how the
width of surfaces scales with the size of surfaces
\cite{KPZ86,KK89,halpin-healy95,laessig98,kotrla98,chin99}.  Recently,
the coupling between surface growth and other degrees of freedom has
been considered in the context of ordering phenomena on growing and
fluctuating surfaces.  For example, the field coupled to the surface
can be different species of deposited
particles\cite{kotrla98,kotrla97,drossel00} or surface reconstruction
order parameters\cite{chin01}.  Unlike equilibrium surfaces, where the
coupling between the surface height degrees of freedom and the
additional field is irrelevant at large length scales\cite{dennijs85},
we found that the coupling between the surface and the reconstruction
orders can not be ignored\cite{chin01}.  The surface and the
additional field are usually coupled topologically.  For example, the
domain walls of order parameters become trapped at the hilltops or in
the valleys on surfaces in 1+1 D
system\cite{kotrla98,kotrla97,drossel00} or on the ridge lines on
surfaces in a 2+1 D system\cite{chin01}.

The fluctuations of the additional degrees of freedom (expressed by
the dynamics of the ``domain walls'') thus become slaved to the
fluctuations of the surface. From the surface perspective, the
additional field is usually irrelevant and passive.  Therefore, part
of the dynamics of the surface and its topological features can be
evaluated from the fluctuations of such domain walls that are pinned
or trapped to them.  We can put probes on the surface to follow these
features dynamically.  In this paper, passive random walkers(PRWs)
are designed to follow hilltops or valley bottoms on the surface.

One of the first applications of the PRWs is to provide a direct way
to measure the dynamical exponent of the surface fluctuations.  We
will also present the phenomena of coalescence of passive random
walkers on Kardar-Parisi-Zhang(KPZ)\cite{KPZ86,KK89} type growth.  The
coalescence of passive random walkers uncovers a river-like network
structure of the surface in space-time.

Our passive random walkers are similar to the passive scalars in
turbulence.  This is not surprising because the KPZ equation is
equivalent to Burgers equation, which describes a fluid\cite{KPZ86}
with different random stirring forces from those in related turbulence
studies.  It is instructive to study the dynamics of the PRWs from
both perspectives.

This paper is organized as follows.  The passive random walkers model
is defined in Sec.\ref{sec:model}.  The application to directly
measuring the dynamical exponent is presented in Sec.\ref{sec:dyns}.
Then we discuss the coalescence of the PRWs in Sec.\ref{sec:clsc}.
The sign of the coupling between the PRWs and the surface is important
for the coalescence phenomena as discussed in Sec.\ref{sec:negK}.  In
Sec.  \ref{sec:RNPS}, we show the river-like network in the space-time
structure of KPZ-type surfaces and the relation between the passive
random walkers and the passive scalars in turbulence is briefly
discussed.  Other possible applications of the PRW model are pointed
out, together with a summary of the results in Sec.\ref{sec:cncls}.

\section{Passive Random Walkers on a Growing Surface}
\label{sec:model}

We couple a PRW to the growth dynamics of a fluctuating surface.  The
movement of the PRW is determined by the local slope of the growing
surface.  The well-known Kim-Kosterlitz (KK) model\cite{KK89} for
surface growth is used as an example.  In the KK model, a surface is
specified by integer height variables on a square lattice.  A single
particle is deposited at a randomly chosen site $i$ if the restricted
solid-on-solid condition ($|\delta h_{ij}| = |h_j - h_i| \leq 1$ for
all nearest neighbor pairs $\langle ij \rangle$) remains satisfied,
otherwise, the deposition of the particle is rejected.  The surface
grows and the stationary state is rough.  The position of the PRW is
updated after $N_{\rm s}^2/v_{\rm g}$ Monte Carlo steps (with $v_{\rm
  g}$ the growth velocity and $N_{\rm s}$ the total number of site),
i.e., when on average one layer of surface material is added,
according to the following rules.  Let $i$ be the position of the
walker.  If there is any neighbor site $j$ for which $h_j > h_i$, then
we move the walker to site $j$.  If there is more than one site higher
than site $i$, then we let the PRW move to either one with equal
probability.  We apply periodic boundary conditions for both the
surface and the PRW motion.

\begin{figure}[p]
  \centering
  \includegraphics[height=7cm]{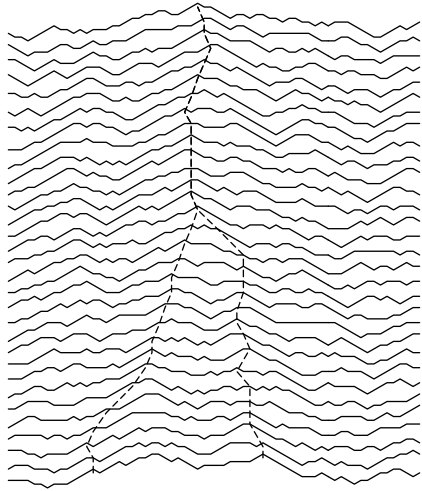}
  \caption{Passive walkers trapped on local hilltops and global
    maximum in 1+1 D simulations.  The dashed lines are the world
    lines of two PRWs and the solid lines are the surface profiles
    during growth.  The passive walkers follow the local maximums
    initially and ultimately merge into each other.  The walkers reach
    the global maximum at a later time. }
  \label{fig:WKS}
\end{figure}

It is known that the continuum limit of the KK model is governed by
the so-called KPZ equation\cite{KK89},
\begin{equation}
  \label{eq:KPZ}
  \frac{\partial h}{\partial t} = 
  \nu\frac{\partial^2 h}{\partial {\bf x}^2} + 
  \lambda\left(\frac{\partial h}{\partial {\bf x}}\right)^2 + \eta_{\rm s},
\end{equation}
with uncorrelated noise,
\begin{equation}
 \langle \eta_{\rm
  s}({\bf x},t)\eta_{\rm s}({\bf x}^\prime, t^\prime) \rangle = D \delta({\bf x},{\bf x}^\prime)\delta(t,t^\prime).
\end{equation}
In the same limit.  the equation of motion of the PRW is
\begin{equation}
  \label{eq:PRW}
  \frac{\partial u}{\partial t} = 
  \kappa \frac{\partial}{\partial {\bf x}} h({\bf x},t)|_{{\bf x}={\bf
  u}(t)},
\end{equation}
where ${\bf u}(t)$ is the coordinate of the PRW at time $t$, and
$\kappa$ is the coupling strength between the PRW and the surface.  In
our lattice model, $\kappa$ is of the same order as $\lambda$.  The
equations of motion, Eqs.(\ref{eq:KPZ})-(\ref{eq:PRW}), imply that the
PRW moves upwards for $\kappa > 0$.  The PRW becomes trapped on a
local maximum of the surface profile as seen in Fig.(\ref{fig:WKS})
for positive $\kappa$.  Once the PRW is trapped on a hilltop, it
follows the motion of the hilltop, which is subject to the surface
fluctuations.  When $\kappa$ is negative, the PRW moves downwards
instead of upwards, and the PRW becomes trapped in a valley bottom
instead of a hilltop.  The relative sign between $\kappa$ and
$\lambda$ affects the dynamics of the PRWs.  We will focus on positive
$\kappa$ for now, and discuss negative $\kappa$ in Sec.\ref{sec:negK}.

\vspace{1cm}

\section{Dynamical Exponent of Passive Random Walkers}
\label{sec:dyns}

One important observation of the dynamics of an upwards moving PRW on
KPZ-type surfaces is that the PRW not only reaches a local maximum in
a relatively short time, but also that in finite-size systems it
ultimately ends up trapped at the {\em global} maximum.  We will
discuss the detailed mechanism of this phenomenon later in
Sec.\ref{sec:clsc}.  Once the PRW is trapped on the global maximum, it
performs a correlated random walk following the global surface
fluctuations.  The surface fluctuates critically and is self-affine.
Namely, the profile of the surface is invariant under the following
rescaling,
\begin{equation}
  \label{eq:scaling}
  x \rightarrow b^{-1}x,\,\,h \rightarrow b^{-\alpha}h,\,\,
  t \rightarrow b^{-1/z_{\rm s}}t,
\end{equation}
where $b$ is a scaling factor, $\alpha$ is the roughness exponent, and
$z_{\rm s}$ is the dynamical exponent of the surface.  The PRW is
slaved to the fluctuations of the interface, therefore we expect
similar dynamical scaling for the world line of the PRW.

The displacement of the PRW, $\Delta {\bf u}(t) = {\bf u}(t) - {\bf
  u}(0)$, should be defined carefully to avoid confusion in
finite-size systems with periodic boundary conditions, where the PRW
moves on a ring in 1D and on a torus in 2D.  In such manifolds, the
winding number of the PRW should be taken into account.  The component
of the displacement, $\Delta {\bf u}(t)$, in direction ${\rm
  \hat{e}_i}$ is defined as $\Delta u_{\rm \hat{e}_i}(t) = n^{+}_{\rm
  \hat{e}_i} - n^{-}_{\rm \hat{e}_i}$, with the actual number of right
moves, $n^{+}_{\rm \hat{e}_i}$, and the actual number of left moves,
$n^{-}_{\rm \hat{e}_i}$ in direction $\hat{e}_i$.

In our numerical simulations, we let the system evolve until both the
surface and the PRW motion reach stationary states. Then the value
of the displacement $\Delta {\bf u}(t)$ is measured as $\Delta u(t) =
\left(\sum_{\rm i} \left(n^{+}_{\rm \hat{e}_i} - n^{-}_{\rm
      \hat{e}_i}\right)^2\right)^{1/2}$.  To accelerate the
simulation, we also adopt a rejection-free algorithm.  Details of this
algorithm can be found in ref.\cite{chin01}.  In this algorithm, time
is counted in terms of the numbers of layers of particles deposited
instead of the conventional Monte Carlo time unit.  We established
earlier that the unit of time in this rejection free algorithm is
linearly proportional to the Monte Carlo time unit\cite{chin01}(see
also \cite{drossel00}).

If the PRW indeed is slaved to follow the global surface fluctuations,
which are invariant statistically under the transformation
eq.(\ref{eq:scaling}), the average distance of the displacement,
$\Delta u(t)$, must obey the scaling form,
\begin{equation}
  \label{eq:scaling2}
  \langle \Delta u(t) \rangle \sim L^\chi 
  {\cal G} \left(\frac{t}{L^{z_{\rm s}}}\right),
\end{equation}
where ${\cal G}(\tau)$ is a universal scaling function.  When $t \ll
L^{z_{\rm s}}$, the hopping events of the PRW are correlated due to
the surface fluctuations, and $\Delta u(t) \sim t^{z_{\rm w}}$, where
$z_{\rm w}$ is the dynamical exponent for the PRW.  At time scales $t
\gg L^{z_{\rm s}}$, the surface fluctuations are limited by the finite
size of the lattice and the PRW becomes like a free uncorrelated
random walker.  Thus, the scaling function ${\cal G}(\tau)$ has the
asymptotic forms, ${\cal G}(\tau)\rightarrow \tau^{1/z_{\rm w}}$ for
$\tau \ll 1$ and ${\cal G}(\tau) \rightarrow \tau^{1/2}$ for $\tau \gg
1$.  The value of the exponent $\chi$ in Eq.(\ref{eq:scaling2})
follows from the fact that $u(t)$ must be independent of $L$ in the
thermal dynamics limit.  Consequently, $\chi$, $z_{\rm w}$, and $z_{\rm
  s}$ are not independent and satisfy the relation $\chi =
\frac{z_{\rm s}}{z_{\rm w}}$.

\begin{figure}[p]
  \centering
  \includegraphics[width=8cm]{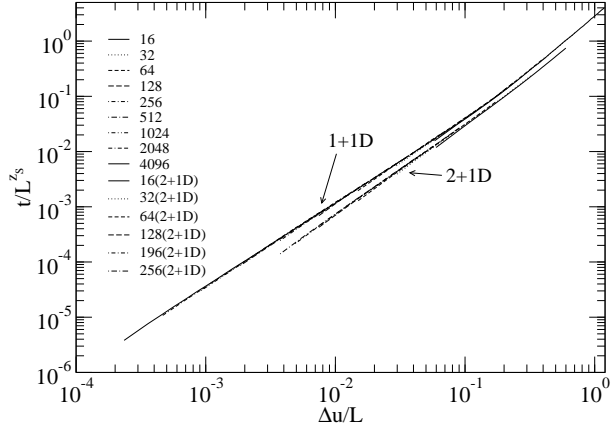}  
  \caption{Collapses of $t$ versus $\Delta u$. We use $z_{\rm s}=1.5$
  for 1+1D, $z_{\rm s}=1.6$ for 2+1D and $\chi=1$ for both cases.}
  \label{fig:clps}
\end{figure}

The dynamical exponent of the walker is not necessarily the same as
the dynamical exponent of the surface.  There is no {\em a priori}
reason for $\chi=1$, i.e., for $z_{\rm w}=z_{\rm s}$.  The numerical
results (Fig.~\ref{fig:clps}) indicate that $z_{\rm w}=z_{\rm s}$, but
why?  $\chi=1$ reflects that there is no length scale other than the
system size $L$ involved in the dynamics of the PRW.  This is in
contrast to free uncorrelated random walks, in which case another
length scale proportional to $t^{1/2}$ will emerge.  The absence of a
new independent length scale is consistent with our picture that the
PRW is typically trapped on a global maximum, hence the fluctuations
of the PRW follow the dynamical scaling of the surface.

\begin{figure}[p]
  \centering
  \includegraphics[width=8cm]{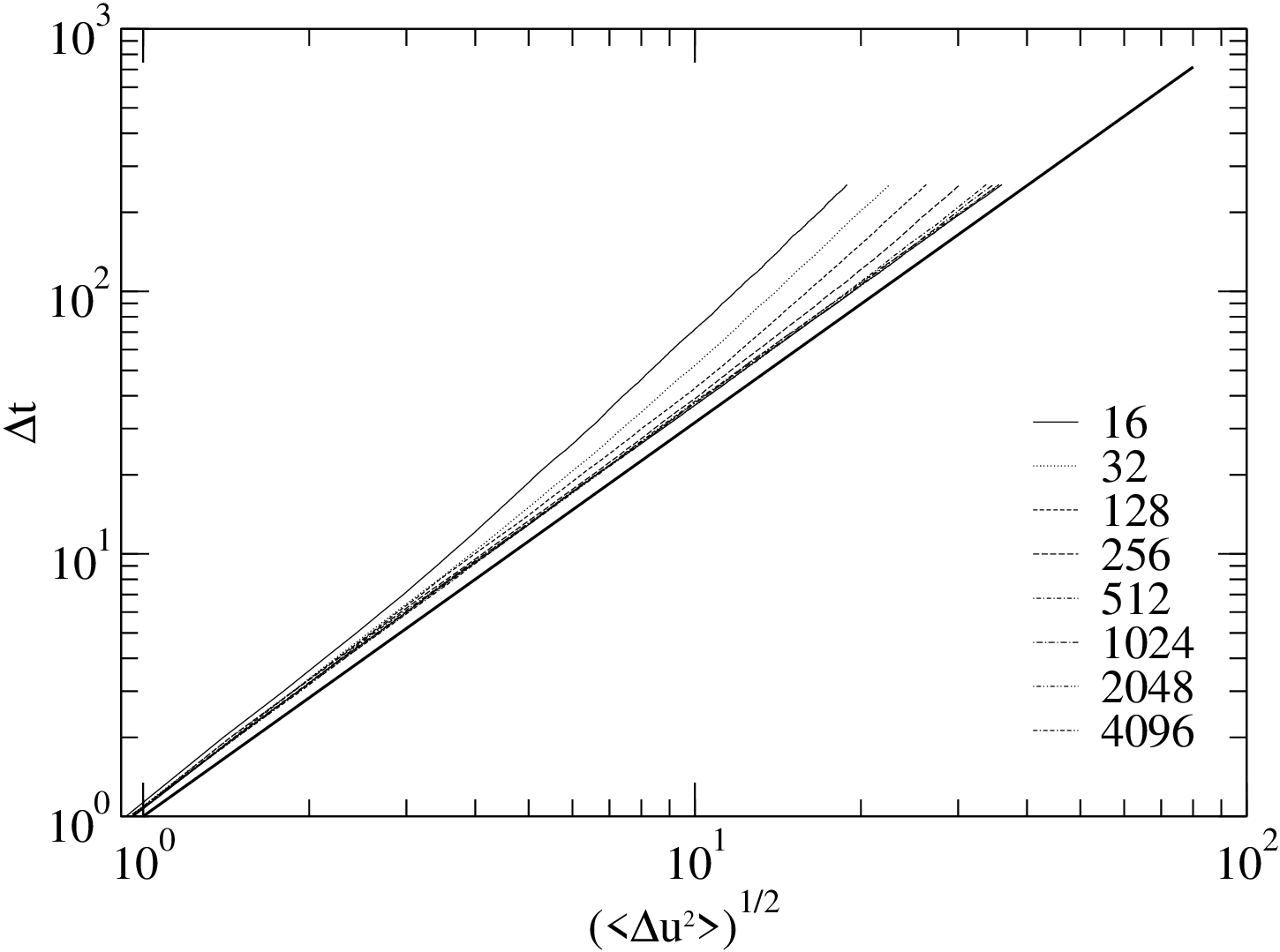}
  \includegraphics[width=8cm]{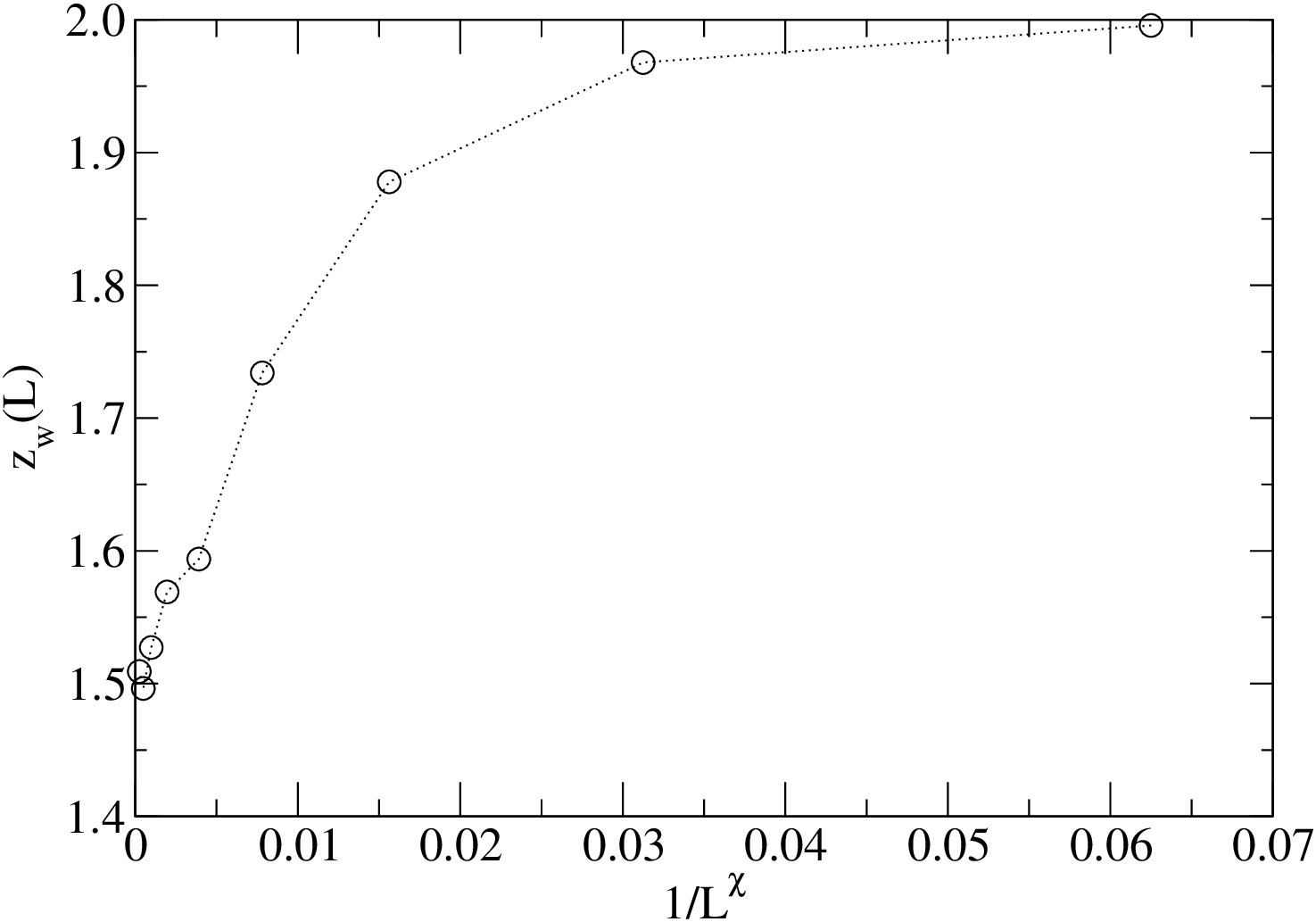}
  \caption{(a)The log-log plot of $t$ versus $\Delta u$ in the 1+1D KK
    model.  For reference, we plot the solid line with a slope of 1.5.
    (b)Fitting of $z_{\rm w}$ for different system sizes in 1+1D.
    It shows the crossover from free uncorrelated random walker
    behavior ($z=2$) to a KPZ-type PRW behavior ($z=1.5$).  The dotted
    line is a convenience to guide the eyes.}
  \label{fig:dxdt1d}
\end{figure}

\begin{figure}[p]
  \centering
  \includegraphics[width=8cm]{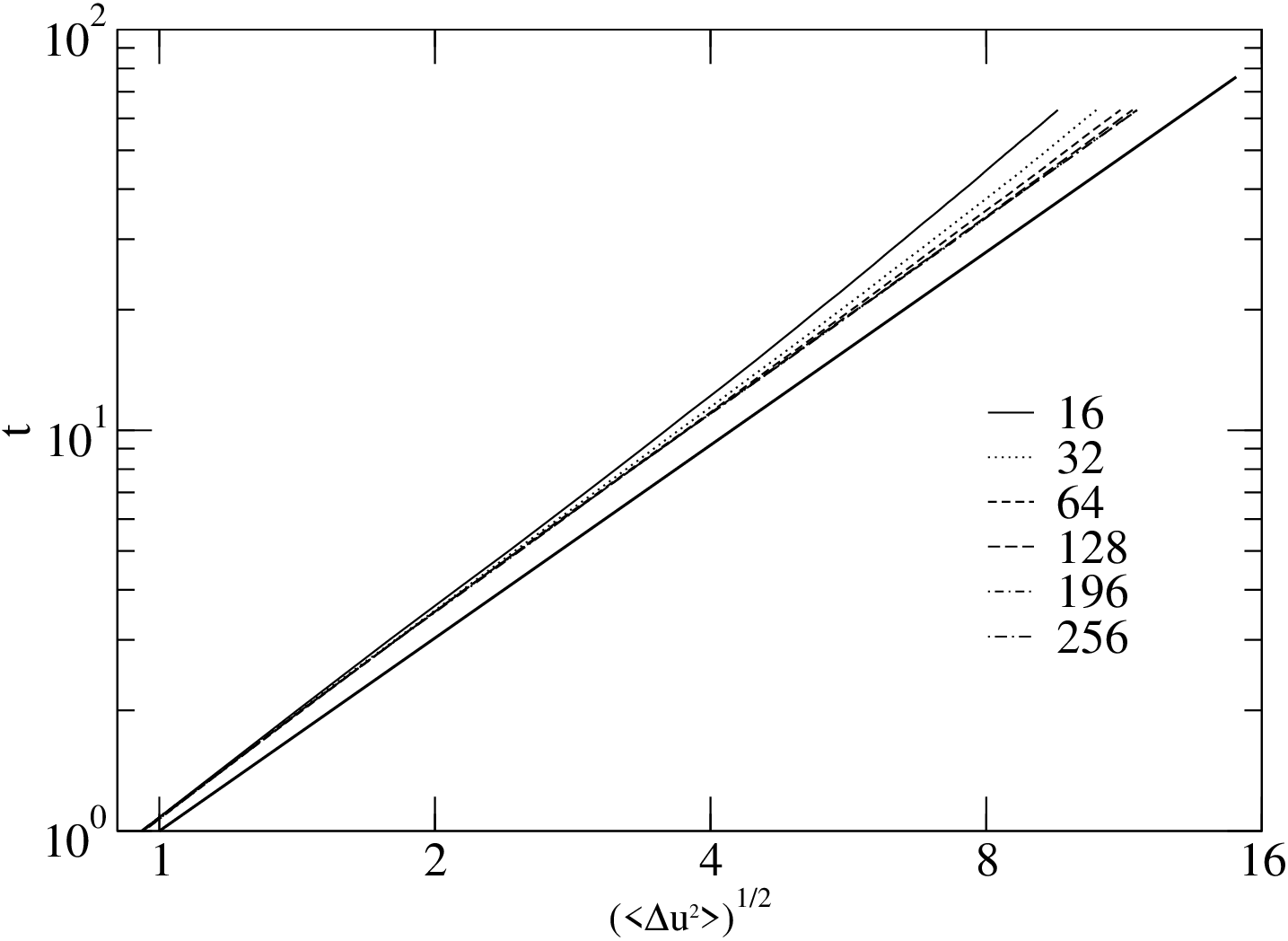}
  \includegraphics[width=8cm]{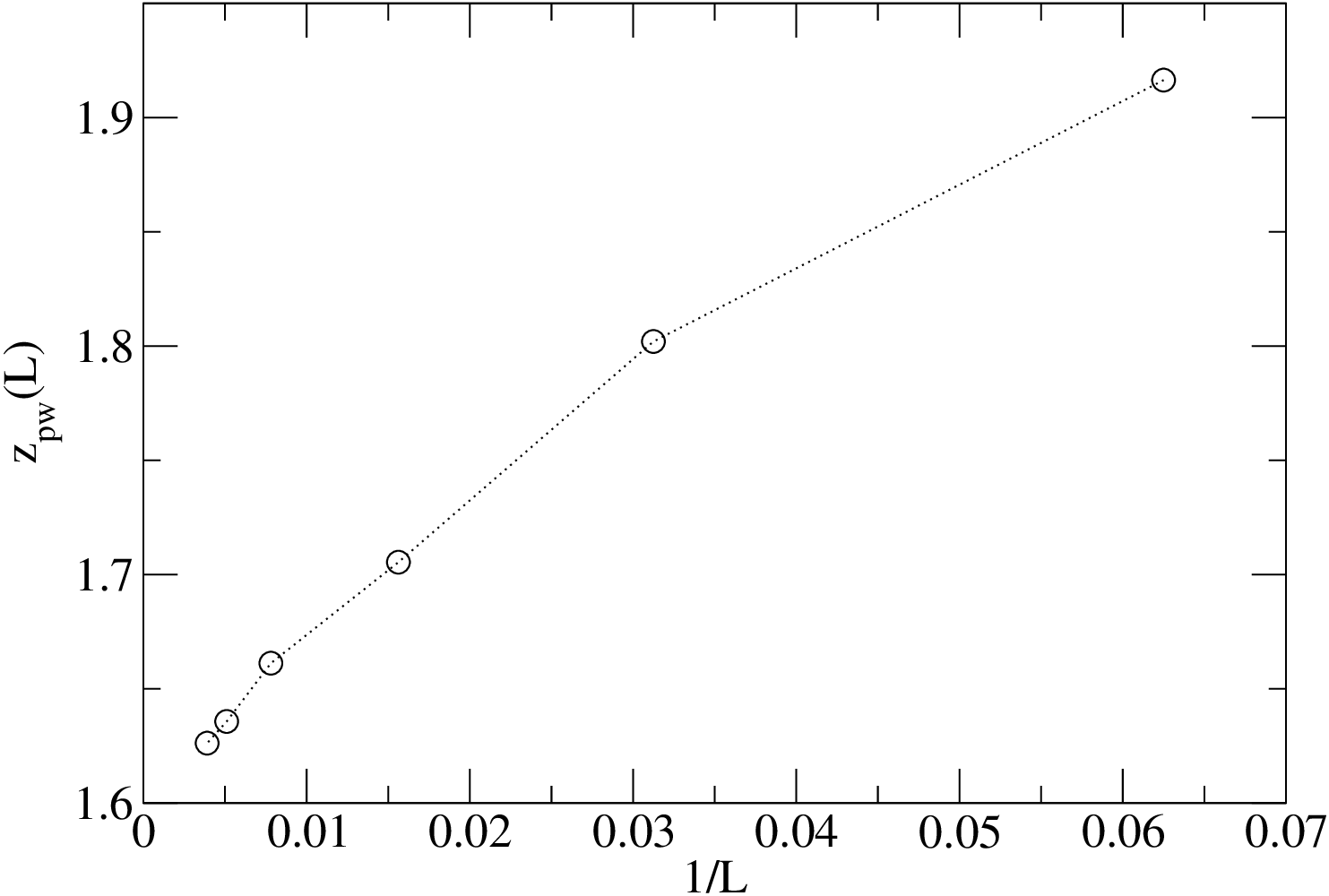}
  \caption{(a)The log-log plot of $t$ versus $\Delta u$ in the 2+1D KK
    model.  For reference, we plot the solid line with a slope of 1.6.
    (b)Fitting of $z_{\rm w}$ for different system sizes in 2+1D.
    It shows the crossover from free uncorrelated random walker
    behavior ($z=2$) to a KPZ-type PRW behavior ($z=1.6$).  The dotted
    line is a convenience to guide the eyes.}
  \label{fig:dxdt2d}
\end{figure}

The main numerical results are shown in
Fig.(\ref{fig:clps})-(\ref{fig:dxdt2d}).  They are obtained by
averaging over about $10^5$ samples in 1+1D and $4\times10^5$ samples
in 2+1D.  The displacements of the PRW are determined up to $t = 256$
in 1+1D and $t = 128$ in 2+1D.  In 1+1D, $\langle\Delta u\rangle$ is
in the order of $10^2$ lattice units at $t=256$.  In 2+1D,
$\langle\Delta u\rangle \sim 10$ at $t = 128$.  Although
$\langle\Delta u\rangle$ is relatively small in 2+1D, the scaling
behaviors are already clear.  Small displacements reflect merely a
small prefactor in the scaling form eq.(\ref{eq:scaling2}).  This is
analogous to a small diffusion constant (which slows the diffusive
process, but does not affect the scaling properties even at small
spatial scales).  Moreover, the asymptotic region of interest is at
small $\tau$ and at small $\Delta u/L$, so the scaling behaviors that
we focus on are in the small $\Delta u$ interval.  By increasing
system sizes, we can extend the scaling region of the scaling
variable, $\Delta u/L$, about two decades.

We check the scaling form eq.(\ref{eq:scaling2}) by collapsing the
data (see Fig.(\ref{fig:clps})) with $z_{\rm s} = 1.5$ in 1+1 D and
$z_{\rm s} = 1.6$ in 2+1 D.  The prediction, $\chi=1$, gives excellent
data collapse, except for the nonuniversal part around $\Delta u
\lesssim 10$ in 1+1D and $\Delta u \lesssim 4$ in 2+1D.  This
nonuniversal behavior is likely caused by the discrete lattice
spacing.  A more detailed fit of the dynamical exponent $z_{\rm w}$
must avoid this small $\Delta u$ region while keeping $\Delta u/L$
small.  In Fig.(\ref{fig:dxdt1d}b)(\ref{fig:dxdt2d}b), we fit $z_{\rm
  w}$ by omitting the data from $\Delta u \lesssim 10$ in 1+1D and
$\Delta u \lesssim 4$ in 2+1D.  It might appear that by ignoring data
from these regions, a large fraction of information in the data is
lost.  In fact, only less than a quarter of the data points fall in
these regions.  The crossover from uncorrelated random walker behavior
($z=2$, at large $\Delta u/L$) to correlated behavior($z\not=2$, at
small $\Delta u/L$) is visible in
Fig.(\ref{fig:dxdt1d}b)(\ref{fig:dxdt2d}b), where the exponent
decreases from 2 to $z_{\rm s}$ as the system size increases.  We
conclude that $z_{\rm w} = 1.50(1)$ in the 1+1D and $z_{\rm w} =
1.60(1)$ in the 2+1D KK model.

Our measurement of $z_{\rm w}$ provides one of the very few {\em
  independent} and {\em direct} measurements of the dynamical exponent
$z_{\rm s}$ on {\em stationary-state} KPZ-type surfaces.  Conventional
approaches determine the dynamical exponent indirectly, through the
measurement of the scaling behaviors of the width of surfaces in the
transient states, starting with a flat initial condition.  The surface
width defined as $W = \langle (h - {\bar h})^2 \rangle^{1/2}$ scales
as $W\sim t^{\beta}$ for $t \ll L^{z_{\rm s}}$.  Then the dynamical
exponent is found by using the scaling relation $z_{\rm
  s}=\alpha/\beta$.  The other method is to measure the correlation
function $g(x,t)= \langle\left(h(x_0+\Delta x, t_0+\Delta t) -
  h(x_0,t_0)\right)^2\rangle \sim (\Delta t)^{2\beta}$ for $\Delta
t<(\Delta x)^{z_{\rm s}}$ at stationary states.  Both methods obtain
the dynamical exponent indirectly through the growth exponent $\beta$.
By tracing the path of the PRW, we are able to provide an independent
and direct measurement of the dynamical exponent without knowledge
about the global width of the surface.

Compared to the conventional methods of measuring the dynamical
exponent, we only need to acquire information of the local slopes
around the PRW.  Thus, it is instructive to understand the connection
between the slope-slope correlations along the path of the PRW and the
global roughening dynamics of the surface.

The displacement of the PRW and the correlation of the slopes are
connected by the fluctuation-dissipation theory.  The scaling of the
dispersion in the PRW displacement can be evaluated from the
slope-slope correlation function along the world line of the
PRW\cite{BOHR93}, i.e.,
\begin{equation}
  \label{eq:FDT}
   \langle ({\bf u}(t) - {\bf u}(0))^2 \rangle  \sim  t \int_{0}^{t} d\tau\,\Phi(\tau)
  \sim t^{1/z_{\rm w}},
\end{equation}
with $\Phi(\tau)$, 
\begin{equation}
   \Phi(\tau)  =  \left\langle \frac{\partial {\bf u}(\tau)}{\partial t}
  \frac{\partial {\bf u}(0)}{\partial t} \right\rangle \sim  \left\langle
  \frac{\partial}{\partial x} h({\bf u}(\tau),\tau)
  \frac{\partial}{\partial x} h({\bf u}(0),0) \right\rangle,
\end{equation}
This correlation function $\Phi(\tau)$ is different from the
conventional slope-slope correlations, ${\tilde \Phi}(r,t)=\langle
\partial_x h(r,t) \partial_x h(0,0) \rangle$, where the correlations
are calculated at specific fixed $r$.  Instead, we need to evaluate
the correlation ${\tilde \Phi}(r,t)$ along the correlated path, the
world line of the PRW, $r=u(t)$.
  
The most general scaling form for ${\tilde \Phi}(r,t)$ is,
\begin{equation}
  \label{eq:PhiScaling}
  {\tilde \Phi}(r,t) = b^{2\eta}{\tilde \Phi} \left( b^{-1}r,
  b^{-z_{\rm s}}t \right),
\end{equation}
where $\eta$ is an exponent yet to be determined.  This scaling form
is equivalent to
\begin{equation}
  \label{eq:PhiScaling2}
  {\tilde \Phi}(r,t)\sim t^{2\eta/z_{\rm s}}F\left(r^{-z_{\rm s}} t \right),
\end{equation}
with $F(y)$ a scaling function.  Since a path of the PRW is determined
by the surface fluctuations, averaging over different paths can be
approximated by averaging over different realizations of the surface
(like different Monte Carlo runs).  $\Phi(\tau)$ is approximately
equal to the value of ${\tilde\Phi(r,t)}$ at $r=\Delta u(t)\sim
t^{1/z_{\rm w}}$.  Thus, $\Phi(\tau)$ scales as
\begin{equation}
  \label{eq:PhiScaling3}
  \Phi(\tau)\sim {\tilde\Phi}(\tau^{1/z_{\rm w}}, \tau) \sim
  \tau^{2\eta/z_{\rm s}} 
  F \left( \tau^{1 - \chi} \right).
\end{equation}

If $\chi \geq 1$, asymptotically, the scaling function behaves as
$F\left( \tau^{1 - \chi} \right) \sim F(0) + F^\prime(0)
\tau^{1-\chi}$, which approaches a constant, for $\tau\rightarrow
\infty$. Hence the correlation function scales as $\Phi(\tau)\sim
\tau^{2\eta/z_{\rm s}}$. Together with Eq. (\ref{eq:FDT}), we obtain
$2+2\eta/z_{\rm s} = 2/z_{\rm w}$.  This relation allows us to
determine $\eta$, which is the scaling dimension of the slope
operator, from the measurement of $z_{\rm w}$.  Furthermore, if
$\chi=1$, then $\eta = \alpha - 1$, provided that the KPZ scaling
relation $\alpha + z_{\rm s} = 2$ holds.  This result, $\eta = \alpha
- 1 $, is consistent with L\"assig's operator production expansion
scheme \cite{laessig98} and our previous work\cite{chin99} for KPZ
type surfaces. Namely, all slope--slope correlations can be obtained by
the naive power counting and the slope operator, $\frac{\partial
  h}{\partial x}$, scales as $x^{\alpha - 1}$.

Our numerical results suggest that $\chi$ is indeed equal to 1 for
upwards moving PRWs.  In 1+1D, renormalization group(RG) calculations
gave ${\tilde \Phi}(r,t)\sim t^{-1/z_{\rm s}}F(r^z/t)$\cite{FNS77}, so
our results imply the value of $z$ is fixed as $z_{\rm w} = z_{\rm
  s}=3/2$ in 1+1 D.  In 2+1 D KPZ-type surface growth, RG
calculations indicated that the fixed point is at strong coupling such
that analytical results can not be obtained
perturbatively\cite{KPZ86}.  To author's knowledge, no analytic
result for the scaling from of ${\tilde \Phi}(r,t)$ has been obtained
for 2+1 D.  Our numerical results support $\eta = \alpha-1$ for
KPZ-type surfaces.

\section{Coalescence of Passive random walkers}
\label{sec:clsc}

Why does an upwards moving PRW typically end up on the global maximum?
To further understand the mechanism of this phenomenon, we study the
dynamics of multiple PRWs on KPZ-type surfaces.  If all upward
moving PRWs eventually reach the global maximum on the finite-sized
KPZ-type surface, then all of them will converge to the same hilltop
even though they start at different positions.  This is true only if
the PRWs have the property of coalescence, namely two PRWs close to
each other merge into each other and move together afterward during
the process of moving up to the global maximum.

Although, in the definition of the dynamics of the PRWs, the PRWs
always move upward, this simple coupling between the PRW and the
surface only guarantees that the PRW reaches a {\em local} maximum,
{\em not} the global one.  However, we find that in KPZ dynamics with
negative $\lambda$ the upwards moving PRWs coalesce.  Two PRWs
separated over a distance $r$ at time $t=0$ merge into each other
after a typical time period $t\sim r^{z_{\rm w}}$ as shown in
Fig.(\ref{fig:WKS}) ( see also (\ref{fig:RN}a) ).  On the contrary,
downwards moving PRWs do not coalesce in $\lambda < 0$ KPZ-type
dynamics and PRWs do not coalesce in Edwards-Wilkinson(EW) type growth
(the $\lambda=0$ point of KPZ equation).

Consider the PRWs sitting on two nearby hilltops that have
similar heights.  The region between the two hilltops is a small and
shallow valley.  If both the hilltops and the valley are relatively
higher than the other parts of the surface, the valley is likely to be
filled up and vanish in a short time.  These two hilltops merge and
the merging event causes the PRWs to coalesce afterward as shown in
Fig.(\ref{fig:WKS}).  The coalescence of the PRWs is likely caused by
merging of two nearby hilltops by filling the valley in between.

This trivial argument seems to imply that PRWs will coalesce for any
type of growing surface as long as the PRWs are trapped on hilltops.
This is not true.  We found that the PRWs on EW type surfaces do not
coalesce, although KPZ-type surfaces and the EW-type surfaces have
the same stationary state and roughness exponent $\alpha=1/2$ in 1+1
D.  The following discussion and the detail analysis in
Sec.\ref{sec:NoiselessKPZ} show how the nonlinear term,
$\lambda\left(\partial_x h\right)^2$, in the KPZ
equation(eq.(\ref{eq:KPZ})) affects the phenomenon of coalescence.

\begin{figure}[p]
  \centering \includegraphics[height=6.5cm,width=7.5cm]{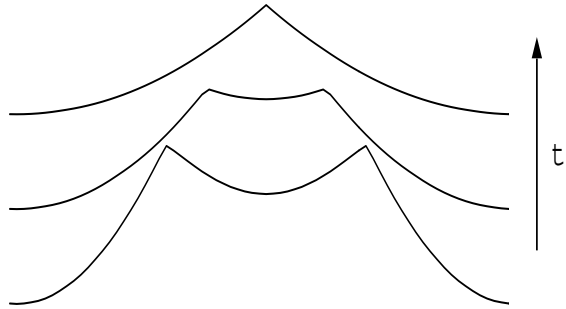}
  \caption{A close look at the coalescence of two hilltops driven by
    the noiseless KPZ equation. The curves in this figure have been
    shifted for clarity.  The two hilltops have a common contact
    point at the bottom of the valley between them.}
  \label{fig:twoHillTops}
\end{figure}

The nonlinear term, $\lambda (\partial_x h)^2$, in the KPZ equation
controls how the growth rate depends on local slopes.  Consider a
plateau on a surface, for positive $\lambda$, the growth rate on the
slope parts of the plateau is larger than the flat top. The flat top
of the plateau expands by lateral growth on the slopes.  In general,
hilltops become flatter than valleys for positive $\lambda$.  Negative
$\lambda$ has an opposite effect, the slope parts grows slower than
the flat tops hence the flat tops become sharper.  Shrinking of
plateau flat tops is an indication of a negative $\lambda$ in the KPZ
equation.  The merging events of two hilltops is equivalent to
shrinking of flat top plateaus.  For example, at larger length scales,
the shallow valley between two nearby hilltops and the hilltops
themselves as a whole acts like a plateau on the surface
(Fig.(\ref{fig:twoHillTops})).  Before the coalescence of the
hilltops, the size of the flat top of the plateau is the distance
between these two hilltops.  While these two hilltops are moving
towards each other by surface fluctuations, the size of the flat
plateau top effectively shrinks, which is an indication of a negative
$\lambda$.  In the KK model, $\lambda$ is negative and the origin of
negative $\lambda$ is due to the restricted solid-on-solid condition
$|\Delta h_{ij}| \leq 1$, which reduces the growth rate to zero at the
regions where the density of $\Delta h = 1$(or $\Delta h=-1$) steps is
1.  With negative $\lambda$ for the KK model, which belongs to the KPZ
universality class, both the hilltops and the PRWs coalesce.

\section{Coalescence of Hilltops in the Noiseless KPZ Equation}
\label{sec:NoiselessKPZ}

We can gain insights about the coalescence of the PRWs on KPZ-type
surfaces by investigating the solution for the noiseless KPZ equation
in the limit $\nu\rightarrow 0$.  In that limit, the evolution of
surfaces with given initial conditions has an analytical solution,
which allows us to study the details of the dynamics of the hilltops.

A detailed derivation of the solution for the noiseless KPZ equation
can be found in \cite{woyczynski98}.  We only present a brief review
of the derivation.  With the Hopf-Cole transformation,
$\phi(x,t)=\exp\left(\lambda h(x,t) /\nu\right)$, we can transform the
noiseless KPZ equation into a simple linear diffusion equation for
$\phi(x,t)$,
\begin{equation}
  \label{eq:diff}
  \frac{\partial \phi}{\partial t} = \nu
  \frac{\partial^2\phi}{\partial x^2}.
\end{equation}
The solution is a superposition of Gaussian functions weighted by the
initial conditions,
\begin{equation}
  \phi(x,t) = \left(2\pi\nu t\right)^{-\frac{1}{2}} \int
  dy\,\phi(y,0)\exp\left(-\frac{(x-y)^2}{\nu t}\right),
\end{equation}
where $\phi(y,0) = \exp\left(\lambda h(y,0)/\nu \right)$ is the
initial condition, determined by the initial height profile $h(y,0)$.
In the limit $\nu \rightarrow 0$, we can simplify the solution
by introducing the velocity, $v = \partial_x h$, as an auxiliary
variable,
\begin{equation}
  v = \frac{\nu}{\lambda}\frac{\partial \ln\Phi}{\partial x} = \frac{ \int dy\,
  \frac{2(x-y)}{\lambda t}
  \exp\left(-\frac{1}{\nu}\left(\frac{(x-y)^2}{t} + \lambda h(y,0)\right)\right)}
{\int dy \exp\left(-\frac{1}{\nu}\left(\frac{(x-y)^2}{t} +
  \lambda h(y,0)\right)\right)}.
\end{equation}
By the method of steepest descent in the limit $\nu\rightarrow 0$,
this yields
\begin{equation}
  v(x,t) =  \frac{-2(x - y_{\rm min}(x,t))}{\lambda t},
\end{equation}
where
$y_{\rm min}$ is a function of both $x$ and $t$, and it is determined
by taking $y$ where $\frac{(x-y)^2}{t} - \lambda
  h(y,0)$ is a minimum for fixed $x$ and fixed $t$.  Namely,
\begin{equation}
y_{\rm min}(x,t)  =  \arg\min_{y} \left(\frac{(x-y)^2}{t} -
  \lambda h(y,0)\right).
\end{equation}
The $y_{\rm min}(x,t)$ are usually only piecewise continuous functions
of $x$, particularly for random initial conditions generated by random
walkers.  Shock waves of the velocity field are formed where $y_{\rm
  min}(x,t)$ is discontinuous.

Assuming $y_{\rm min}(x,t)$ is constant between $x$ and $y_{\rm min}$,
we can further integrate $v(x,t)$:
\begin{equation}
  \label{eq:sltnh2}
  h(x,t) = \frac{-(x - y_{\rm min}(x,t))^2}{\lambda t} + c_{\rm h},
\end{equation}
where $c_{\rm h}$ is the integration constant.  The solution for
$h(x,t)$ is reduced to the problem of finding $y_{\rm min}$ and $c_{\rm
  h}$.  

\begin{figure}[p]
  \centering
  \includegraphics[height=6cm,width=7.5cm]{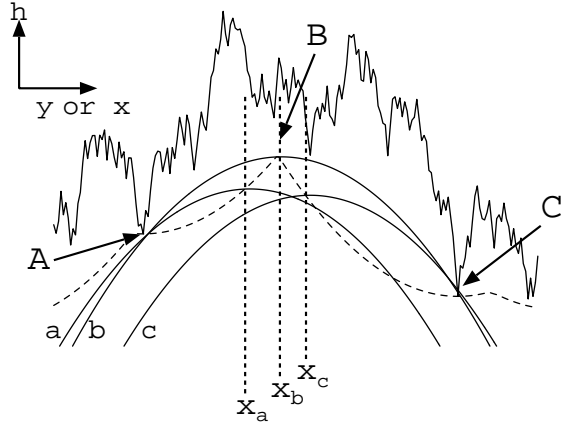}
  \caption{ The parabolic shape probe tip scans through point
    $x_a$,$x_b$, and $x_c$ from left to right.  Curves (a),(b),and (c)
    are parabola which are centered at $x_a$, $x_b$, and $x_c$
    respectively. Curve (a)/(b) only contact with the initial random
    surface at point A/B.  Curve (b) contacts with the initial surface
    at both point A and C.  While the probe is scanning through $x_b$,
    the first contact point jumps from A to C. Thus, $y_{\rm min}(x)$
    is discontinued at point $x_b$ and the path of the tip creates a
    cusp at point C. The dashed curve is the surface generated by the
    path of the probe tip.}
  \label{fig:tsfm1}
\end{figure}

Determining $y_{\rm min}$ for $\lambda < 0$ is equivalent to finding
$y$ for which the vertical distance is the shortest between the
initial height configuration, $h(y,0)$, and the parabola,
$\frac{(x-y)^2}{\lambda t}$, centered at $x$, while the parabola is
approaching the surface from below.  $y_{\rm min}$ is the first
contact point between the two curves $h(y,0)$ and
$\Psi(y;x,t)=\frac{(x-y)^2}{\lambda t} + c$ while $c$ is adjusted to
$c(x,t) = h(y_{\rm min},0) - \frac{(x-y_{\rm min})^2}{\lambda t}$
(Fig.(\ref{fig:tsfm1})).  The contact point $y_{\rm min}$ is where
$h(y,0) - \frac{(x-y)^2}{\lambda t} $ is a minimum.  Once $y_{\rm
  min}$ is found, $h(x,t)$ follows from eq.(\ref{eq:sltnh2}) with
$c_{\rm h}=h(y_{\rm min},t)$. Thus, we have $h(x,t) = \Psi(x;x,t) =
c(x,t)$, which is the top of the parabola.  The surface between $x$
and $y_{\rm min}(x,t)$ is also parabolic in shape at time $t$.  It is
possible that there exists more than one first contact point for
certain $x$ and $t$.  If this happens, $y_{\rm min}(x,t)$ as a
function of $x$ will be discontinuous at that specific $x$.

This solution has an intuitive graphical interpretation, which is
useful for understanding the coalescence phenomena.  The evolution of
the surfaces is generated by a sequence of geometry transformations on
the surfaces at an earlier time.  For negative $\lambda$, the new
surface at time $t$ is obtained by scanning the original surface from
below with a probe that has a parabolic shaped tip, $\Psi(y; x,t) =
\frac{(x-y)^2}{\lambda t} + c(x,t)$(Fig.(\ref{fig:tsfm1})) for
negative $\lambda$.  (For positive $\lambda$, the probe scans the
surface from above instead.)  The probe is centered at position $x$,
and $c(x,t)$ (the vertical position of the top of the tip) is adjusted
by moving the probe up and down.  When the probe approaches the
surface from below, it stops when it starts to come into contact with
the surface.  The horizontal position of the first contact point
between the probe and the surface is $y_{\rm min}(x,t)$. Since the
probe stops moving upward once it comes in contact with the surface,
we might think of the vertical position of the tip of the probe,
$c(x,t)$, as the height of the new surface seen by the probe at $x$ at
time $t$.  The probe scans through the surface and plots a new surface
that gives the surface evolved by the noiseless KPZ equation at a
later time.  Because the parabolic shape of the probe, the new surface
usually consists of a collection of parabola shaped segments(the
dashed curves shown in Fig.(\ref{fig:tsfm1})).

With this geometrical interpretation of the exact solution of the
noiseless KPZ equation, we are able to visualize how the singular hill
tops are formed and to discuss the dynamics of coalescence of the hill
tops. If the local average curvature over a region such as between A
and C in Fig.(\ref{fig:tsfm1}) is larger than the curvature of the tip
of the probe at time $t$, then the tip cannot reach the surface
everywhere between point A and C.  When the probe scans through this
region, at point B in Fig.(\ref{fig:tsfm1}), there are two first
contact points for the parabola, and $y_{\rm min}(x)$ is discontinued
at this point.  At points where the function $y_{\rm min}(x)$ is not
continuous, the path of the tip forms cusps, which is are hilltops of
the new surface.  The position of the hilltop in Fig.\ref{fig:tsfm1}
is determined by the positions of the two contact points A, C and the
curvature of the probe tip.

\begin{figure}[p]
  \centering \includegraphics[height=6.5cm,width=7.5cm]{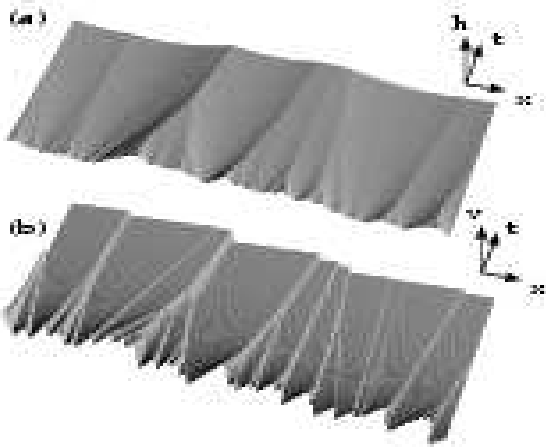}
  \caption{(a)The evolution of the surface under transformation
    eq.(\ref{eq:tsfm}). Hilltops coalesce and form tree-like
    structures. (b) The coalescence of shock waves in the velocity
    field.  The velocity is defined as $v=\frac{\partial h}{\partial
      x}$.}
  \label{fig:coal1}
\end{figure}

We can also write the solution of $h(x,t)$ as a functional
transformation $\cal T$ defined as
\begin{equation}
\label{eq:tsfm}
{\cal T}_t(h(x,t_0)) \equiv  h(x,t+t_0) = \min_y\left( h(y,t_0) -
\frac{(x-y)^2}{\lambda t}\right).
\end{equation}
This transformation has the following property,
\begin{equation}
  {\cal T}_{t_1+t_2}(h(x,t_0))={\cal T}_{t_2}{\cal T}_{t_1}(h(x,t_0)).
\end{equation}
The geometrical interpretation of the solution in this form is similar
to the Huygens principle growth algorithm\cite{tang90}, except that
parabola shaped wave fronts instead of circular or spherical wave
fronts are used.  The transformation eq.(\ref{eq:tsfm}) can be applied
iteratively and is suitable for evaluating the evolution of surfaces
numerically.

We evaluate the evolution of a random initial surface with
Eq.(\ref{eq:tsfm}).  The coalescence of the hilltops is shown clearly
in Fig.(\ref{fig:coal1}a).  Why do those hilltops coalesce and how do
they move?  Will hilltop B in Fig.(\ref{fig:tsfm1}) move towards
point A or point C?  Answers to these questions can be addressed
with the equation of motion of the hilltops.

The curvature of the probe at time $t$ is equal to $1/(\lambda t)$.
The probe becomes broader and broader with time.  As long as the
absolute value of the local curvature around the contact points, A or
C, is much larger than the absolute value of the curvature of the
probe, the positions of these two contact points do not change.  In
this case, the height difference $ h(x_{\rm A},t) - h(x_{\rm C},t)$,
where $x_{\rm A}$ and $x_{\rm C}$ are the horizontal coordinates of
points A and C respectively, is also a constant.  It can be written as
\begin{eqnarray*}
  & & h(x_{\rm A},t) - h(x_{\rm C},t) = \\
  & &(h(x_{\rm A},t) -
  h(x_{\rm B},t))-(h(x_{\rm C},t)-h(x_{\rm B},t)). 
\end{eqnarray*}
The curve between A and B and the curve between B and C of the new
surface are two parabolas.  With $h(x_{\rm A},t) - h(x_{\rm B},t) =
(x_{\rm A} - x_{\rm B})^2/(\lambda t)$ and $h(x_{\rm C},t) - h(x_{\rm
  B},t) = (x_{\rm C} - x_{\rm B})^2/(\lambda t)$, we obtain the equation
of motion of the hilltop,
\begin{equation}
  \frac{\partial x_{\rm B}}{\partial t}  = 
   t^{-1}\left( x_{\rm B} - \frac{x_{\rm A} + x_{\rm C}}{2}\right).
\end{equation}
This equation implies that the hilltops always move toward the
closest of the two contact points.  Moreover, the closest contact
point is also the higher one.  If two neighboring hilltops share the
same contact point, and that contact point is the highest one for
both hilltops, then these two hilltops will merge into one around
the position of the shared contact point as illustrated in
Fig.(\ref{fig:twoHillTops}).  A sequence of coalescence events of the
hilltops forms a tree-like structure of world lines as in Fig
(\ref{fig:coal1}a).

The dynamics of the coalescence of the hilltops in the noiseless KPZ
equation is totally deterministic and only depends on initial
surfaces.  In the noisy case, the dynamics becomes stochastic,
however, the qualitative behavior of the coalescence of the hilltops
is still sustained even when the surface is perturbed with randomly
depositions.  The nonlinear term causes the coalescence of the hill
tops and remains essential even when the dynamics is stochastic.  To
check that the nonlinear term is responsible for the coalescence in
the stochastic dynamics, we apply our upwards moving PRW model to
EW-type surfaces.  The upwards moving PRWs for EW type surface do not
coalesce although the density of the PRWs is slightly higher on the
hilltops.

\section{downwards moving passive random walkers}
\label{sec:negK}
So far we have focused only on the case in which $\kappa$ and $\lambda$
have opposite signs, i.e., when the PRWs move upward($\kappa > 0$) in
the KK model ($\lambda<0$).  The discussion in the preceding section
suggests that the shape of the surface has very different
characteristics between a hilltop and a valley bottom on the surface
with noiseless KPZ dynamics. Namely, hilltops are sharp with
discontinued slopes while valley bottoms are rounded and with
continuous slopes.  The symmetry is broken by the nonlinear term.  We
also found that valley bottoms vanish by themselves, instead of
coalescing with others valleys.

If $\kappa$ is negative, the PRWs move toward the valley bottoms.  It
is interesting to see how the PRWs respond to such qualitative aspects
of the surface in the presence of stochastic noise. In the
deterministic case, the hilltops and valley bottoms intertwine, the
number of valley bottoms and the number of hilltops both decrease.
This is not true for the stochastic case.  New microscopic hilltops
and valley bottoms are being formed constantly by the noisy
depositions of particles.  Is the coalescence of the PRWs stable under
such perturbations?  Of upwards moving PRWs in the KK model,
coalescences are stable against the noise.  As we will see next, the
coalescence of the downwards moving PRWs is not stable against random
depositions.

Consider the upwards moving PRWs. Suppose noise splits the hilltops by
creating a small valley between them.  This small valley is unstable
against the dynamics as we have seen in the preceding section.  These
two hilltops will merge again soon as a result of the nonlinear term
in the KPZ equation. The nonlinear KPZ dynamics stabilizes the
coalescence of the upwards moving PRWs.

Next, imagine several downward moving PRWs moving into the same
valley.  The aggregation of the PRWs is not stable against the noisy
perturbation.  Depositions of particles on the valley split the
valley.  A PRW on the original valley bottom can go to either one of
those two valleys with equal probability, so both valleys will be
occupied by PRWs.  Do the new created subvalleys always merge back
into each other?  A valley bottom does not move in the noiseless
case; they will remain separated by the hilltop.  The hilltop does not
vanish unless it is driven away to merge with other hilltops, which is
a much less likely process than the splitting.  Therefore, dynamics
like this does not stabilize the aggregation of downwards moving PRWs.
They tend to be separated under the growth dynamics and we expect that
the dynamics of downwards moving PRWs, trapped in valley bottoms, is
dominated by the perturbation of random deposition rather than the
deterministic nonlinear term in the dynamics.  In Fig.(\ref{fig:RN}c),
we show a typical simulation for downwards moving PRWs.  Instead of
coalescence, we find that the aggregation clusters of PRWs are not
stable and split constantly.

Coalescence of the upwards moving PRWs for negative $\lambda$ is an
important feature of KPZ-type surfaces.  Not only does it ensure that
the PRWs end up trapped on the global maximum; it also reveals the
space-time structure of KPZ-type surfaces in detail, i.e., hilltops
(valley bottoms) coalesce as time evolving for negative (positive)
$\lambda$.

Our analysis on the KPZ equation applies to the phenomena of coalescence
of the shock waves in the inviscid($\nu\rightarrow 0$) noiseless
Burgers equation.  The velocity in the Burgers equation is the slope on
the KPZ-type surfaces, i.e., $v=\partial_x h$.  By differentiation of
eq. (\ref{eq:KPZ}), it becomes
\begin{equation}
  \label{eq:Burgers}
  \frac{\partial v}{\partial t} = \lambda v\cdot\frac{\partial
  v}{\partial x} + \frac{\partial^2 v}{\partial x^2} + \eta_{\rm B}.
\end{equation}
The noise term satisfies $\langle \eta_{\rm B}(x,t)\eta_{\rm B}(0,0)
\rangle \sim \partial_x^2 \delta(x)\delta(t)$.  Because the slopes of
the surface are discontinuous at the hilltops, the hilltops on
KPZ-type surfaces correspond to the shock waves in the Burgers
equation.  The coalescence of the hilltops of the surface indicates
that the shock waves coalesce, as illustrated in
Fig.(\ref{fig:coal1}(b)).  Coalescences of shock waves described by
the noiseless Burgers equation have been reported in a numerical study
of inelastic collisions of particles in 1+1
dimensions\cite{ben-naim99}.  Bohr and Pikovsky also showed that the
zeros in the velocity field of the Burgers equation
coalesce\cite{BOHR93}.  The above analysis for the coalescence of KPZ
hilltops translates directly into the coalescence of shock waves in
the Burgers equation with noise.  Our analysis on coalescences of PRWs
on a growing surface presented here not only provides further
intuitive and detailed understanding of the mechanism of these
coalescence phenomena but also explains, in both deterministic and
stochastic dynamics, the different behaviors between hilltops and
valleys, which are distinct types of zeros in the velocity field in
the Burgers turbulences.

\section{River-like network and Passive Scalars}
\label{sec:RNPS}

\begin{figure}[p]
  \centering
  \includegraphics[width=7.5cm]{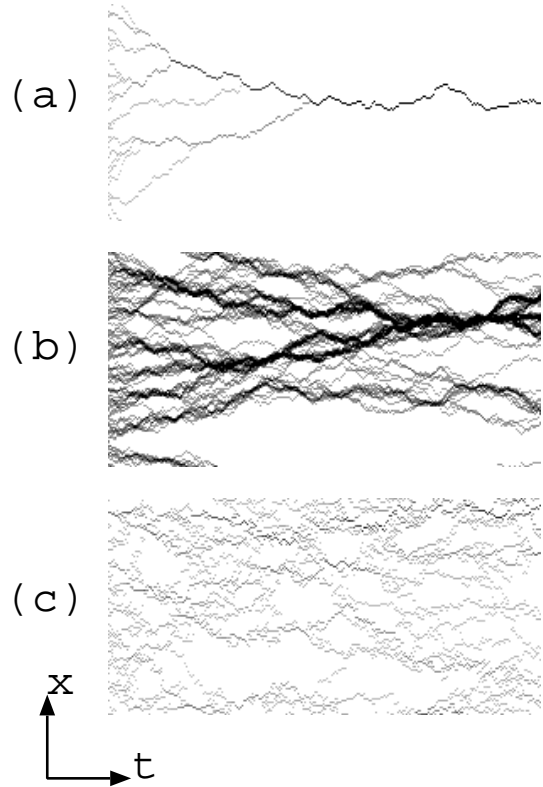}
  \caption{World lines of passive walkers for different parameters. The
    lines are the paths of 128 PRW on a system of size L=128. The
    darkness is proportional to the number of PRWs for each site.  In
    (a), the system is simulated in the limit $l_{\rm
      c}=\left(\frac{D_{\rm pw}}{K}\right)^{\frac{1}{2-z}}\rightarrow
    0$ for upward moving PRWs.  In (b), we choose $l_{\rm c} \sim L$,
    the system size.  (c) is the result of downwards moving PRWs in the
    limit $l_{\rm c}\rightarrow 0$.  Even without a diffusion term in
    eq.~(\ref{eq:PRW2}), the downwards moving PRWs do not coalesce.
    The dynamics of the downwards moving PRWs on KK-type surface is
    dominated by the uncorrelated random depositions.}  
  \label{fig:RN}
\end{figure}

To further understand the stability of the aggregations of PRWs, we
study how random forces, in addition to the driving forces from
surface slopes, are applied to the PRWs directly and change the tree-like
structure of the PRW world lines in Fig.(\ref{fig:RN}a).  Is the
tree-like structure stable?  Random noise modifies the equation of
motion of the PRW as
\begin{equation}
  \label{eq:PRW2}
  \frac{\partial u}{\partial t} = 
  \kappa \frac{\partial}{\partial x} h(x,t)|_{x=u(t)} + \eta_{\rm w}(x,t),
\end{equation}
with $\eta_{\rm w}$ the uncorrelated noise, $\langle \eta_{\rm w}
(x,t)\eta_{\rm w}(0,0) \rangle = D_{\rm w}\delta(x)\delta(t)$. $D_{\rm w}$ is
the diffusion constant of the PRW if $\kappa = 0$.

First consider a single PRW on a surface.  The second term on the
right hand side of eq.(\ref{eq:PRW2}) introduces the ordinary
diffusive behavior, i.e., $\Delta u(t) \sim (D_{\rm w}t)^{1/2}$.  The
scaling behavior due to the first term, $\Delta u(t)\sim (K_{\rm w}
t)^{1/z_{\rm w}}$, ($K_{\rm w}$ is a constant determined by $\kappa$
and $\lambda$) still dominates the large scale behavior since
$1/z_{\rm w} > 2$. However, the PRW appears as diffusive instead of
super-diffusive at length scales smaller than $l_{\rm c} \sim \left(
  \frac{D_{\rm w}}{K_{\rm w}} \right)^\frac{1}{2-z_{\rm w}}$.

The diffusive noise also affects the coalescence of multiple PRWs.
For small perturbations, the aggregation is stable because the PRWs
cannot escape from the hilltop by diffusion if the hilltop is high
and large enough.  The net effect of diffusion simply broadens the
coalescence cluster.  However, if we increase the strength of the
random noise, the PRWs can escape from the hilltop and be caught by
other hilltops nearby with finite probability.  In this case, the
cluster of coalescence PRWs splits and the world lines of the PRWs
form a braid-like network instead of a tree structure.  A typical
configuration of such a braid-like network is shown in
Fig.(\ref{fig:RN}b).  Both the tree-like and the braid-like network
resemble river-like networks of different length scales in
Nature\cite{fract_river_basin}.
    
The crossover scale between tree-like networks and braid-like networks
is given by $l_{\rm c}$. We observe diffusive behaviors locally at
scales less than $l_{\rm c}$.  The braid-like networks emerge at the
crossover scale, $l_{\rm c}$, and the tree-like structures are recovered
at scales much larger than $l_{\rm c}$.

Furthermore, we can consider the evolution of the probability
distribution function, $P(x,t)$, of finding a particle at position $x$
at time $t$.  Note that the total number of PRWs is conserved in our
model.  Therefore, $P(x,t)$ has the standard conserved form,
\begin{equation}
  \frac{\partial P(x,t)}{\partial t} = \nabla \cdot j(x,t),
\end{equation}
where $j(x,t)$ is the PRW current.  The current is the sum of the advected
part and the diffusive contribution, i.e., $j(x,t) = \kappa v P(x,t) + D_{\rm
  w}\partial_x P(x,t)$.  Hence, we have
\begin{equation}
  \frac{\partial P(x,t)}{\partial t} = \kappa \nabla (v P(x,t)) +
  D_{\rm w}\nabla^2 P(x,t).
\end{equation}
This is simply the equation of motion of passive scalars, $P(x,t)$, in
a fluid.  In the limit $D_{\rm w}\rightarrow 0$, the passive scalars
will be concentrated at a globe hilltop as we have seen in the
previous sections if $\kappa$ and $\lambda$ have opposite sign for a
finite system.

\section{Concluding remarks}
\label{sec:cncls}
Motivated by the domain wall dynamics on nonequilibrium fluctuating
surfaces, we study the dynamics of passive random walkers on KPZ-type
growing surfaces.  We show that, although the coupling rule between a
PRW and a surface is defined locally, the PRW typically reaches the
maximum of the surface over a distance $x$ in a time scale $t\sim
x^{{\rm z}_s}$ and ``feels'' the fluctuations of the surface over the
same length scale.  Two PRWs separated by $\Delta x$ coalesce after
$\Delta t \sim (\Delta x)^{{\rm z}_s}$.  The fluctuations of the
positions of such PRWs follow the same dynamical scaling as the KPZ
fluctuations.  We verify this scaling behavior numerically.  Tracing
the paths of the PRWs on KPZ-type surfaces is an effective method to
measure the dynamical exponent directly in the stationary state.

In addition to the scaling, the dynamics of passive random walkers
reveals detailed specific space-time structures of a KPZ-type growing
surface, i.e., hilltops coalesce with each other and the world lines
of the PRWs form self-affine tree-like structures.  We provide an
analytical argument explaining this phenomenon based on the noiseless
Burgers-KPZ equation.  We show how the nonlinear term $(\partial_x
h)^2$ is responsible for this nontrivial coalescence phenomenon in both
the noiseless and noisy case.  The noiseless KPZ equation also allows 
us to understand why downwards moving PRWs do not coalesce, due to the
asymmetry between hilltops and valley bottoms.

We want to emphasize that the effectively attractive interactions
which makes the PRWs coalesce are not only topological but also robust
against perturbations.  For KPZ-type growth, the $(\partial_x h)^2$
term breaks the symmetry between hilltops and valley bottoms during
growth and introduces nontrivial dynamics of the hilltops for negative
$\lambda$ and the valley bottoms for positive $\lambda$.  The
singularities on KPZ-type surfaces generated by the nonlinear terms
stabilize the aggregations of the PRWs against other perturbations.
The coalescence of PRWs reveals one of the important features of the
complicated space-time structure of the surface.  An interesting
extension of this study is to see how to generalize this to different
universality classes of surface dynamics.  We already point out that
in EW growth, because of the particle-hole symmetry, PRWs do not
coalesce.  For other nonlinear models, in which the nonlinear terms
break up-down symmetry, PRWs may be driven in other nontrivial ways
and the surface may show interesting space-time structures.

Depending on the strength of the noise applied to the PRWs, the world
lines of PRWs form tree-like or braid-like networks in space-time.
The tree-like or the braid-like networks resemble river type networks
in nature.  The braid-like networks are the result of two competing
mechanisms, namely, the coalescence between PRWs caused by the
fluctuations of surfaces and the diffusion of PRWs caused by directly
applied noise.  These two mechanisms define a crossover length scale
$l_{\rm c}$, and give rise to the braid-like networks at this scale.
One might expect to see similar crossover phenomena in the study of
passive scalars in Burgers turbulence.

Previous studies on fluctuating nonequilibrium surfaces were focused on
the scaling behavior.  This study shows that, beyond global scaling
laws, such as the scaling of the global interface width, nonlinear
surface growth dynamics leads to intriguing detailed aspects of their
space-time structures.  We hope studying these different perspectives
of nonequilibrium dynamics can lead to a better understanding of these
complex systems.

This research is supported by the National Science Foundation under
grant No. DMR-9985806.  The author thanks Marcel den Nijs for many
useful discussions and critical reading of the manuscript.


\end{document}